\documentclass[dvips]{article}
\usepackage{graphicx}
\newcommand{\doublespacing}{\let\CS=\@currsize\renewcommand{\baselinesstrech}
{2.0}\tiny\CS}
\begin{document}
\newcommand{\bd}{\begin{document}}
\newcommand{\ed}{\end{document}}
\newcommand{\bc}{\begin{center}}
\newcommand{\ec}{\end{center}}
\newcommand{\bfr}{\begin{flushright}}
\newcommand{\efr}{\end{flushright}}
\newcommand{\lt}{\left}
\newcommand{\rt}{\right}
\newcommand{\vs}{\vspace}
\newcommand{\hs}{\hspace}
\newcommand{\beq}{\begin{equation}}
\newcommand{\eeq}{\end{equation}}
\newcommand{\lb}{\linebreak}
\newcommand{\pb}{\pagebreak}
\newcommand{\mb}{\makebox}
\newcommand{\fb}{\framebox}
\newcommand{\mc}{\multicolumn}
\newcommand{\ben}{\begin{enumerate}}
\newcommand{\een}{\end{enumerate}}
\newcommand{\bit}{\begin{itemize}}
\newcommand{\eit}{\end{itemize}}
\newcommand{\oln}{\overline}
\newcommand{\un}{\underline}
\newcommand{\lefq}{\lefteqn}
\newcommand{\ba}{\begin{array}}
\newcommand{\ea}{\end{array}}
\newcommand{\beqa}{\begin{eqnarray}}
\newcommand{\eeqa}{\end{eqnarray}}
\newcommand{\beqas}{\begin{eqnarray*}}
\newcommand{\eeqas}{\end{eqnarray*}}
\newcommand{\bfg}{\begin{figure}}
\newcommand{\efg}{\end{figure}}
\newcommand{\bds}{\begin{displaymath}}
\newcommand{\eds}{\end{displaymath}}
\newcommand{\btb}{\begin{tabbing}}
\newcommand{\etb}{\end{tabbing}}
\newcommand{\para}{\parallel}
\newcommand{\pad}{\partial}
\newcommand{\nn}{\nonumber}
\newcommand{\la}{\leftarrow}
\newcommand{\ra}{\rightarrow}
\newcommand{\lgla}{\longleftarrow}
\newcommand{\lgra}{\longrightarrow}
\newcommand{\La}{\Leftarrow}\newcommand{\Ra}{\Rightarrow}
\newcommand{\Lra}{\Leftrightarrow}
\newcommand{\Lgla}{\Longleftarrow}
\newcommand{\Lgra}{\Longrightarrow}
\newcommand{\lan}{\langle}
\newcommand{\ran}{\rangle}
\renewcommand{\a}{\alpha}
\renewcommand{\b}{\beta}
\newcommand{\g}{\gamma}
\newcommand{\G}{\Gamma}
\renewcommand{\d}{\delta}
\newcommand{\eps}{\epsilon}
\newcommand{\Th}{\Theta}
\newcommand{\s}{\sigma}
\newcommand{\lam}{\lambda}
\newcommand{\D}{\Delta}
\newcommand{\ds}{\displaystyle}
\newcommand{\vare}{E}
\newcommand{\pr}{\prime}
\newcommand{\ro}{\rho}
\newcommand{\nab}{\nabla}
\newcommand{\m}{\mu}
\newcommand{\n}{\nu}
\newcommand{\Sg}{\Sigma}
\newcommand{\p}{\pi}
\newcommand{\R}{I\!\!R}
\newcommand{\om}{\omega}
\newcommand{\Om}{\Omega}
\newcommand{\ovra}{\overrightarrow}
\newcommand{\ze}{\zeta}
\newcommand{\vart}{\vartheta}
\newcommand{\tri}{\triangle}
\newcommand{\f}{\frac}
\newcommand{\iny}{\infty}
\newcommand{\pro}{\propto}
\renewcommand{\arraystretch}{1.25}

\bc {\huge $\kappa$ deformed Dirac equation in crossed magnetic and electric fields} \ec

\vs{1cm}

\bc
{\it D. Nath{\footnote {e-mail : debrajn@gmail.com} \\Department of Mathematics\\
Vivekananda College\\
Kolkata-700063,India.\\

P. Roy{\footnote{e-mail : pinaki@isical.ac.in}}\\
Physics \& Applied Mathematics Unit \\
Indian Statistical Institute \\
Kolkata - 700 108, India.}} \ec
\vs{2.5cm}
\bc {\large {\un{Abstract}}} \ec 
We obtain solutions of the $(2+1)$ dimensional $\kappa$ deformed Dirac equation in the presence of crossed magnetic and electric  fields. It is shown that the $\kappa$ deformed Landau levels are modified in the presence of the electric field. Contraction of Landau levels has also been examined and it has been shown that the contraction depends on a critical magnetic field which is independent of the deformation parameter in first order approximation.
\newpage
\section{Introduction}
Over the years application of various deformed symmetry in the context of quantum physics have been studied by many authors. Out of various deformed symmetries, a particularly interesting one is the quantum deformation of the Poincar\'e algebra or the $\kappa$ deformed Poincar\'e algebra \cite{p1,p2,p3,p4}. Based on this deformed algebra a modified Dirac equation, usually called the $\kappa$ deformed Dirac equation \cite{p2} was constructed. To get an understanding of how the deformation affects observables like energy various quantum mechanical models e.g, the Dirac Coulomb problem \cite{biden}, relativistic Landau problem \cite{pr1}, Aharonov-Bohm interaction \cite{pr2}, anomalous magnetic moments \cite{pr3} have been studied within the framework of $\kappa$ deformed Dirac equation. Very recently the effect of the deformation parameter on the Dirac oscillator problem \cite{edil1,edil2}, integer quantum Hall effect \cite{edil3}, uniformly accelerated observer \cite{hari1}, electrodynamics \cite{hari2} etc. have been examined within the above framework. In this paper our aim is to consider the $\kappa$ Dirac equation in the presence of an interesting interaction, namely crossed magnetic and electric fields. The standard Dirac equation in the presence of this interaction was earlier shown to admit exact solutions \cite{canuto,lam} and here it will be shown that the $\kappa$ deformed Dirac equation with this interaction is solvable too, at least to first order in the deformation parameter. In particular, we shall obtain the spectrum and determine how the $\kappa$ deformed Landau levels \cite{pr1,edil3} are modified due to the presence of the electric field. A comparison of the energy levels with those of the undeformed system has also been made. In this context it may be recalled that relativistic Landau levels in the presence of a crossed electric field show contraction as the electric field strength approaches the magnetic field strength through values less than the later. Here we shall also investigate whether or not the deformation parameter has any effect on such a contraction.

\section{Formalism}
To begin with let us note that the deformed Poincar\'e algebra is given by \cite{p2}
\beq
\ba{l}
[P_i,P_j]=0,~~~~[P_i,P_0]=0,~~~~[M_i,P_j]=i\eps_{ijk}P_k,\\

[M_i,P_0]=0,~~~~[L_i,P_0]=iP_i,~~~~[L_i,P_j]=i\delta_{ij}\eps^{-1}sinh(\eps P_0),\\ 

[M_i,M_j]=i\eps_{ijk}M_k,~~~~[M_i,L_j]=i\eps_{ijk}L_k,~~~~[L_i,L_j]=-i\eps_{ijk}[M_kcosh(\eps P_0)-\frac{1}{4}\eps^2 P_kP_iM_i],
\ea
\eeq
where $P_\mu=(P_0,P_i)$ are the deformed momenta, ($M_i,L_i$) are spatial rotation and boost generators. The quantum group parameter $\eps$ is given by $\eps=\kappa^{-1}=\lim_{R\ra\infty}(Rlnq)$ where $R$ is the de Sitter radius and $q$ is a real deformation parameter. Based on this algebra the $(2+1)$ dimensional $\kappa$ deformed Dirac equation to $O(\eps)$ can be found as \cite{p2,biden}
\beq\label{dirac1}
\left\{\g_0P_0-\g_iP_i+\f{\eps}{2}[\g_0(P_0^2-P_iP_i)-mP_0]\right\}\psi=m\psi,
\eeq
where $\psi=\left(\ba{c}\psi_1 \\ \psi_2\ea\right)$ and the gamma matrices are given in terms of Pauli matrices by \cite{edil1}
\beq
\g_0=\s_z,~~~~\g_1=i\s_y,~~~~\g_2=-is\s_x.
\eeq
and $s=\pm 1$ indicates two spin orientations. In order to incorporate the magnetic and electric fields it is now necessary to gauge Eq.(\ref{dirac1}). It may be noted that as we are dealing with a one body problem no co-multiplication is necessary and also for the above form of the Dirac equation operator ordering ambiguities do not arise \cite{biden}. Thus we gauge Eq.(\ref{dirac1}) in the usual way i.e,
\beq
P_\mu\ra {\hat P}_\mu=P_\mu+A_\mu.
\eeq
and obtain from Eq.(\ref{dirac1})
\beq\label{gauge}
\left\{\g_0{\hat P}_0-\g_i{\hat P}_i+\f{\eps}{2}[\g_0({\hat P}_0^2-{\hat P}_i{\hat P}_i)-m{\hat P}_0]\right\}\psi=m\psi,
\eeq
where 
\beq
{\hat P}_0=P_0-{\cal E}y,~~~~{\hat P}_x=p_x-By,~~~~{\hat P}_y=p_y.
\eeq
Note that the above choice of gauge potentials implies that the direction of the magnetic field is perpendicular to the plane while the electric field is directed along the $y$ axis. Now identifying $P_0=H=E$, Eq.(\ref{gauge}) can be written as
\beq\label{eigen1}
\left[H_0-\f{\eps}{2}({\hat P}_0^2-{\hat P}_i{\hat P}_i-m\s_z{\hat P}_0)\right]\psi=E\psi,
\eeq
where the undeformed Hamiltonian $H_0$ is given by
\beq
H_0=\g_0(\g_i{\hat P}_i+m)-V,~~~~V={\cal E}y.
\eeq
The eigenvalue equation in Eq.(\ref{eigen1}) can be written as
\beq\label{eigen2}
\left[\a\s_x{\hat P}_x+s\a\s_y{\hat P}_y+m_1\s_z-({\cal E}y+E)\right]\psi=0,~~~~\a=1+\f{m\eps}{2}\s_z,~~~~m_1=m-\f{s\eps \hbar B}{2}.
\eeq
It may be noted that due to our choice of the gauge potentials the motion in the $x$ direction is free and consequently we may take the spinor $\psi$ to be of the form
\beq\label{wf2}
\psi(x,y)=e^{ik_xx}\phi(y)=e^{ik_xx}\left(\ba{c}\phi_1\\\phi_2\ea\right),
\eeq
where $k_x$ denotes the momentum in the $x$ direction. Now substituting (\ref{wf2}) in Eq.(\ref{eigen2}) we find
\beq\label{eigen3}
[s\a\s_yp_y+\a\s_x(\hbar k_x-By)+m_1\s_z-({\cal E}y+E)]\phi=0.
\eeq
In the next section we shall find solutions of Eq.(\ref{eigen3}).

\section{Solution}
It is easy to see from Eq.(\ref{eigen3}) that the components $\phi_{1,2}$ are non trivially coupled. In order to disentangle the components we now write
\beq\label{wf1}
\phi=[s\a\s_yp_y+\a\s_x(\hbar k_x-By)+m_1\s_z+({\cal E}y+E)]\chi.
\eeq
Then substituting $\phi$ from Eq.(\ref{wf1}) in Eq.(\ref{eigen3}), it can be shown that the spinor components $\chi_{1,2}$ satisfy the following set of coupled second order equations
\beq\label{intert}
\left(\ba{cc}\displaystyle J+\f{sB}{\hbar} & \displaystyle-\f{s\eps}{\hbar(1-\f{m\eps}{2})}\\\displaystyle\f{s\eps}{\hbar(1+\f{m\eps}{2})} & \displaystyle J-\f{sB}{\hbar}\ea\right)\left(\ba{c}\chi_1\\\chi_2\ea\right)=0,
\eeq
where $J$ is a second order differential operator given by
\beq\label{j1}
J=-\f{d^2}{dy^2}+\f{{\hat\a}^2B^2-{\cal E}^2}{{\hat\a}^2\hbar^2}y^2-2\left(\f{Bk_x}{\hbar}+\f{{\cal E}E}{{\hat\a}^2\hbar^2}\right)y+k_x^2+\f{m_1^2-E^2}{{\hat\a}^2\hbar^2}.
\eeq

The above set of equations can be easily decoupled and we finally obtain the following fourth order differential equations satisfied by $\chi_{1,2}$ :
\beq\label{j}
\ba{l}
\displaystyle\left(J-\f{sB}{\hbar}\right)\left(J+\f{sB}{\hbar}\right)\chi_1=-\f{{\cal E}^2}{{\hat\a}^2\hbar^2}\chi_1,\\

\displaystyle\left(J+\f{sB}{\hbar}\right)\left(J-\f{sB}{\hbar}\right)\chi_2=-\f{{\cal E}^2}{{\hat\a}^2\hbar^2}\chi_2,
\ea
\eeq
where ${\hat\a}=\sqrt{det~\a}$.

It is easy to recognize $J$ as the Hamiltonian of a displaced harmonic oscillator shifted in the energy scale {\it provided}  the magnetic field strength is greater than a critical value, that is, $\displaystyle B>B_c=\f{|{\cal E}|}{\hat\a}$ and in this case if $f_n$ is a harmonic oscillator eigenfunction with eigenvalue $\Om_n$, then
\beq\label{om}
Jf_n=\left[\Om_n+\f{m_1^2}{{\hat\a}^2\hbar^2}-\f{({\hat\a}\hbar k_x\b+E)^2}{{\hat\a}^2\hbar^2(1-\b^2)}\right]f_n,~~n=0,1,2,\cdots,
\eeq
where 
\beq
\ba{l}
\Om_n=\displaystyle\f{1}{{\hat\a}\hbar}\sqrt{{\hat\a}^2B^2-{\cal E}^2}(2n+1),~~~~\displaystyle\b=\f{{\cal E}}{{\hat\a} B},\\\\

\displaystyle f_n=\sqrt{\f{1}{2^nn!}\sqrt{\f{\sqrt{{\hat\a}^2B^2-{\cal E}^2}}{\pi{\hat\a}\hbar}}}H_n\left(\sqrt{\f{\sqrt{{\hat\a}^2B^2-{\cal E}^2}}{{\hat\a}\hbar}}t\right)~exp[-\f{\sqrt{{\hat\a}^2B^2-{\cal E}^2}}{2{\hat\a}\hbar}t^2],~~~~t=y-\f{({\hat\a}^2\hbar k_xB+{\cal E}E)}{({\hat\a}^2B^2-{\cal E}^2)}.
\ea
\eeq
and $H_n(z)$ denotes Hermite polynomials. Then identifying $\chi_{1,n}$ with $f_n$, we find from Eqs.(\ref{j}) and (\ref{om}) 
\beq\label{spec}
(E_{n}+{\hat\a}\hbar k_x\b)^2=m_1^2(1-\b^2)+2{\hat\a}^2\hbar B(1-\b^2)^{\f{3}{2}}n,~~n=0,1,2,\cdots,\\
\eeq
The two independent set of eigenfunctions are given by
\beq
\chi_n=\left(\ba{c}1\\c_+(s)\ea\right)f_n~~or~~\left(\ba{c}1\\c_-(s)\ea\right)f_{n-1},~~~~\displaystyle c_{\pm}(s)=\f{1}{s\b}\sqrt{\f{1-\f{m\eps}{2}}{1+\f{m\eps}{2}}}\left(s\pm\sqrt{1-\b^2}\right).
\eeq

Now from Eq.(\ref{spec}) the energy levels of the original Dirac equation can be found to be 
\beq\label{spec1}
E_{n,\pm}=-{\hat\a}\hbar k_x\b\pm\sqrt{m_1^2(1-\b^2)+2{\hat\a}^2\hbar B(1-\b^2)^{\f{3}{2}}n}
\eeq
From Eq.(\ref{spec1}) it is seen that $E_{n,\pm}$ depends on $\eps$ through the first term on the r.h.s as well as $m_1$ under the radical sign. To separate the $\eps$ dependent part of the spectrum we now expand the r.h.s of the above expression in powers of $\eps$ to obtain
\beq\label{speceps}
\ba{lcl}
E_{n,\pm}&=&-\hbar k_x\b\pm\sqrt{m^2(1-\b^2)+2\hbar B(1-\b^2)^{\f{3}{2}}n}\\
&&\mp\displaystyle\f{sm\hbar B\sqrt{1-\b^2}\eps}{\sqrt{m^2+2\hbar B(1-\b^2)^{\f{1}{2}}n}}+O(\eps^2).
\ea
\eeq

\subsection{Analysis of the spectrum}
The energy values in Eqs.(\ref{spec1}) or (\ref{speceps}) are the $\kappa$ deformed Landau levels in the presence of a homogeneous crossed electric field. It may be noted that presence of the parameter $s$ results in the following change: $c_{\pm}(1)=c_{\mp}(-1)$. As a check on our calculations it may be noticed out that the spectrum in (\ref{spec1}) reduces to the known ones for certain values of the parameters. For example, when $\eps=m=0$ it reduces to the spectrum obtained in ref\cite{bas,peres,nath}. Also, one may recover the $\kappa$ deformed Landau levels \cite{pr1,edil3} when ${\cal E}=0$. As mentioned before the spectrum (\ref{spec1}) is acceptable only when $\b<1$ (or $B>B_c$). It is seen from Eq.(\ref{spec}) that as $\b$ approaches the critical value $1$, the terms including the second term (containing the quantum number $n$) under the radical sign tend to zero and the energy levels tend to a constant value. In other words the Landau levels contract and tend to a constant value as $B\ra B_c$. However, ${\hat\a}=1$ to $O(\eps)$ and consequently the critical magnetic field is independent of the deformation parameter $\kappa$ at this order of approximation. This situation is somewhat similar to ref\cite{biden} where the first order correction to the $\kappa$ deformed hydrogen atom spectrum was found to be vanishing. Finally from (\ref{speceps}) it follows that $E_{n,+}(\eps)-E_{n,+}(\eps=0)<0$ while $E_{n,-}(\eps)-E_{n,-}(\eps=0)>0$ and these two results indicate the net effect of deformation on the energy levels.

Let us now examine the other possibilities regarding the behavior of the Landau levels with respect to the variation of $\b$. We have seen that the Landau levels contract as $B\ra B_c$. When $B=B_c$, the second term on the r.h.s of Eq.(\ref{j1}) vanish and $J$ can be identified with the Hamiltonian of a {\it linear} potential. Although the Schr\"odinger equation for a linear potential admits exact solutions in terms of Airy function, bound states do not exist over the {\it whole} real line \cite{flugge}. Next we consider $B<B_c$. In this case it is seen from Eq.(\ref{j1}) that the coefficient of the second term on the r.h.s becomes negative and consequently $J$ represents the Hamiltonian of a shifted {\it inverted} harmonic oscillator. For the inverted harmonic oscillator there are only discrete complex energy eigenvalues for which bound states do not exist \cite{barton} although this potential is interesting in other contexts e.g, tunneling \cite{sch}. So, we conclude that for $B\leq B_c$ (or $\b\geq 1$) Landau levels do not exist.

\section{Conclusion}
In conclusion we have obtained solutions of the $\kappa$ deformed Dirac equation in crossed magnetic and electric fields to the first order in the deformation parameter. A comparison of the energy levels with and without deformation has also been made. We have also examined in detail contraction of the Landau levels and it has been shown that the critical magnetic field is independent of the deformation parameter to $O(\eps)$. As for future work we feel it would be of interest to search for other interactions solvable within the present framework so that future experiments may provide some estimate on the magnitude of the deformation parameter. Another direction of work may be to solve the Dirac equation keeping terms of order $\eps^2$ and examine dependence of the critical magnetic field on the deformation parameter.

\ed